\documentclass[twocolumn,aps]{revtex4}
\usepackage{graphicx}
\usepackage{subfigure}
\usepackage{amsmath,amsfonts}

\begin{document}
\title{Elastic behavior of spherical nanodroplets in head-on collision}
%
\author{Sangrak Kim}
\email[E-mail address: ]{srkim@kgu.ac.kr}
\affiliation{Department of Physics,\\ Kyonggi University\\ 94-6 Eui-dong, Youngtong-ku, Suwon 440-760, Korea}
\date{\today}
%
\begin{abstract}
Simulation results for head-on collisions of equal-sized spherical polymer nanodroplets using molecular dynamics are presented. Elastic behavior of an initial compressed phase for the colliding droplets is analyzed. Deformations and contact radii of the nanodroplets are compared with the Hertzian model of elastic solid balls. It is found that at least the initial phase of collision can be explained by this continuum model, except at the very moment of the beginning of collision.

\end{abstract}
%
\maketitle
%
\section{introduction}

A nanodroplet has so large a surface-to-volume ratio that its behavior is quite different from the ordinary macroscopic droplets which we see in every-day phenomena. Recently, there are growing interests in nanodroplets collisions\cite{solid}-\cite{ink}, since they are closely related with applications such as nanofluidics, drug delivery, ink-jet printing, \textit{etc}. In particular, Deming and Mason\cite{drug} made nanodroplets for fighting cancer. Their eventual goal is to make possible that the nanodroplets can harmlessly enter into cells and then release medicinal cargo. But making better drug-delivery vehicles still needs a fundamental understanding of nanodroplets behavior.

Contact or collision problems in general are very complicated to model theoretically or even  numerically. Here, we focus on the head-on collisions of  two equal-sized spherical polymer nanodroplets using molecular dynamics. Molecular dynamics (MD) has become an indispensable tool to study the phenomena in nano scale\cite{md}. To understand the initial compressed phase of the collision, deformations, longitudinal forces and contact radii, \textit{etc} are measured from the simulations. The simulation results are analyzed with the Hertzian model\cite{hertz}-\cite{leroy} of colliding elastic solid balls. Originally, the Hertz model deals with compressed elastic balls such as steel balls. It assumes Hooke's law between stress and strain. As far as I know, there is not yet any attempt to connect nanodroplet collision with the Hertzian theory of elasticity. Thus this is a first report to explain the initial phase of droplet collision.

We will first present the Hertzian theory of elasticity for equal-sized spherical elastic balls.  Next, we describe MD simulation method and results. Then, these simulation results are compared with the theoretical predictions. Finally, we will summarize our findings.

\section{Theory}
\begin{figure}
\centering
\begin{tabular}[t]{l l}
a)\raisebox{8.3ex}[0cm][0cm] {\includegraphics[scale=.2]{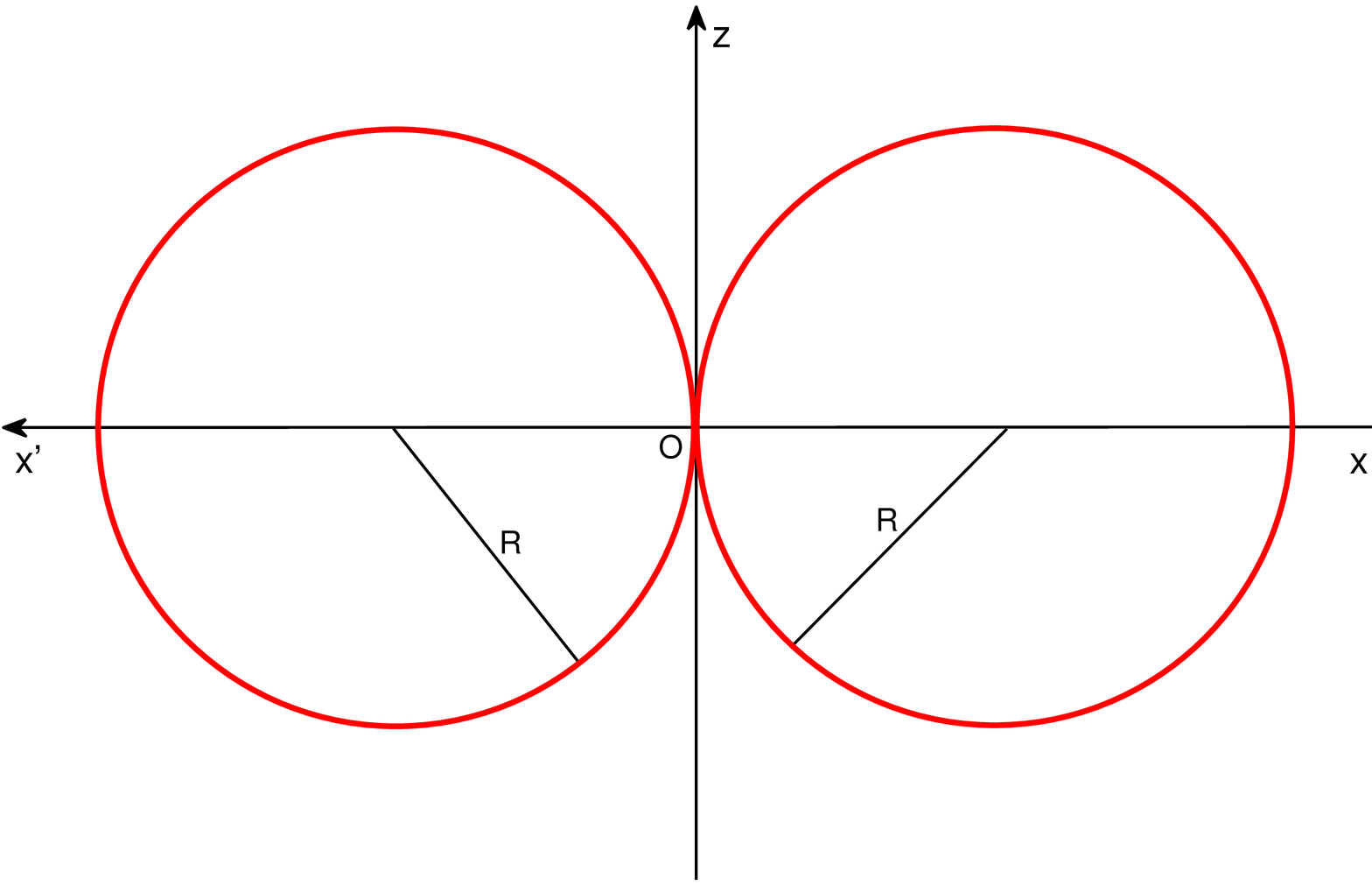}} &
b) \includegraphics[scale=.27, angle=90]{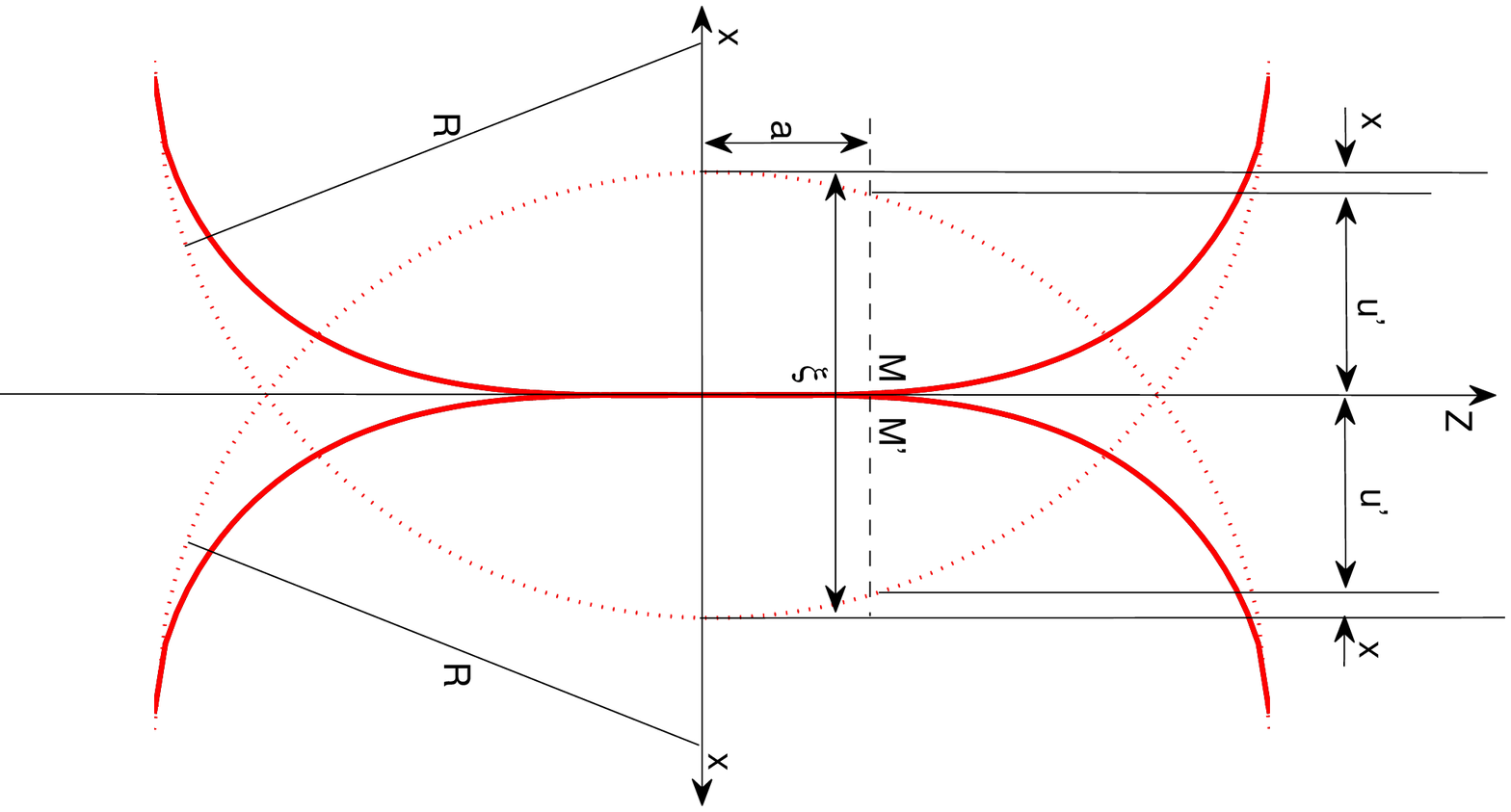}
\end{tabular}
\caption{(Color on-line) a) Two equal-sized spherical elastic balls with a radius $R$ at the moment of contact. b) A compressed state. Dotted lines represent the boundaries of the two balls at the moment of their contact. They are compressed with a contact radius $a$ and total deformation $\xi$. The balls just begin to touch each other at the points M and M', respectively.}
\label{balls}
\end{figure}

When two elastic balls are colliding with each other, conservation laws for energy, linear momentum and angular momentum are  generally applied before and after the collision. In this case, the collision process is treated like a black box: we ignore the collision process itself. However, we want to know what is going on during the collision process. This collision process seems to be similar to that of the statical compression of two balls, which can be explained by the Hertz model. Here, we will summarize the Hertzian model\cite{hertz}-\cite{leroy} of contacting elastic balls which have an equal-sized spherical shape with a radius $R$. The contact is made by pushing them together. Fig. \ref{balls}a) shows the very moment of contact. Let us denote the coordinate system $x$ and $x'$ as being positive in either direction from their contact point. Fig. \ref{balls}b) shows their squeezed state by a force $F_n$ where $F_n$ is a normal component of an applied force from the other ball. The deformed surfaces are thus described by equations of the form $x = x(y, z)$ for the right ball and $x' = x'(y, z)$ for the left ball, respectively, as in Fig. \ref{balls}. Let $u$ and $u'$ be the x-components of the displacement vectors of the contact point on the deformed surfaces of the two balls, respectively. They are symmetrically deformed about an axis connecting the centers of balls. Then, from Fig. \ref{balls}b), we have the equality,
\begin{equation}
    x + u + x' + u' = \xi, \label{condition}
\end{equation}
where $\xi$ is a total deformation of the colliding balls. Thus, the total deformation $\xi$ varies with the force $F_n$.

The material is assumed to be isotropic and homogeneous. Linear elastic Hooke's law holds between stress and strain. It is further assumed that there is no adhesion between balls. Furthermore, we consider only a small $\xi$ limit. In this limit, the total deformation $\xi$ is given by
\begin{equation}
 \label{deformation}
    \xi = {F_n}^{2/3} (\gamma^2 \frac{2}{R})^{1/3},
\end{equation}
where $\gamma \equiv 3 (1-\nu^2) / 2 Y$, $\nu$ is a Poisson's ratio and $Y$ is a Young's modulus. The spherical shapes of balls become flat with the collision as shown in Fig. \ref{balls}b). The contact radius $a$ of this flat contact circle is also given by
\begin{equation}
 \label{c_radius}
    a = {F_n}^{1/3} (\gamma \frac{R}{2})^{1/3}.
\end{equation}

\section{Simulation results}
Now, we simulate head-on collisions of equal-sized spherical polymer nanodroplets. To keep the spherical shape of the droplets, we choose polymer nanodroplets\cite{heine}, since a polymer nanodroplet can have less or no evaporation during the simulation.

First, let us briefly describe the simulation method. The molecular interactions are taken as follows. The polymer chains are modeled by a rather abstract but well-studied bead-spring model\cite{sides}. The chain length of a polymer is $L = 10$. All monomers in the same droplet interact with each other through the Lennard-Jones (LJ) potential, given by
\begin{equation}
    V_{LJ}(r) = 4\epsilon[(r/\sigma)^{-12}-(r/\sigma)^{-6}]. \label{ljpot}
\end{equation}
Neighboring monomers in the same chain in addition interact with the finite extension non-linear elastic (FENE) potential,
\begin{equation}
    V_{FENE}(r) = -\frac{k_F}{2} r_0 log[1-(r/r_0)^{2}], \label{fene}
\end{equation}
where $k_F$ is a spring constant and $r_0$ is a maximum length within which the chain can be maintained. In our simulations, we choose $r_0 = 1.5$ and $k_F = 30.0$. Between molecules in different droplets, only repulsive term of LJ potential is applied so that the droplets repel each other to match with the Hertz model.

We performed MD simulations by using LAMMPS\cite{lammps} package with some modifications or additions, if necessary. Visualizations of the simulation system were done by using VMD\cite{vmd}. Hereafter all the quantities will be expressed in the LJ reduced units.

We prepare two colliding spherical droplets as follows. First, we make a polymer melt with chain length of $L = 10$ such that their positions and velocities correspond to the density of $\rho = 0.5$ and temperature of $T = 0.5$. To make a spherical droplet, we connect each molecule with a virtual harmonic spring with a small spring constant. The other end of each spring is connected to the center of the simulation box to compress the droplet towards its center. Then we can have a spherical droplet with radius $R = 22.87$ with a thermodynamic state of $\rho = 0.83$ and temperature of $T = 0.5$. After the shape of droplet becomes spherical, we turn off the connection of harmonic spring. The number of molecules composing one droplet is $n = 40,000$ so the total number of molecules in the system is $N = 2 n = 80,000$. Another colliding droplet partner is made with the configuration at different time in equilibration process. Thus they are at the same thermodynamic state but with different configurations. Next, they are displaced an appropriate distance . Then the two droplets are approaching each other with the same speed $v$ to collide, so that their relative velocity is $g_n = 2 v$. We choose a significantly larger value of cutoff length $r_c = 5.0$ for LJ interaction, compared to the usual value of  $r_c = 2.5$, since, otherwise, they will experience unwanted forces, when they enter the interaction cutoff region from the far distance.

Since the molecules in different droplets repel each other, the droplets are eventually recoiled. We can measure their respective recoil velocities and thus the restitution coefficient $e$ is determined. The coefficient of (normal) restitution is given by
\begin{equation}
    e = -\frac{g_n'}{g_n}, \label{eqmotion2}
\end{equation}
where $g_n$ and $g_n'$ are normal components of relative velocities of the colliding objects before and after collision, respectively. Anyway, $e$ is related with the forces $F_n$, the material properties and the impact velocity $g_n$, \textit{etc}. The simulation results are summarized in Table \ref{restitution}. Notice that a coefficient of restitution $e$ is strongly dependent on the approaching velocity $v$, since colliding objects are liquid droplets, not the solid clusters. The larger the approaching velocity $v$ is, the smaller a coefficient of restitution $e$ becomes in conformance with the earlier results \cite{kuni}. Their recoil velocities reduce significantly, since their energy is dissipated in the collision. Thus their coefficient of restitution $e$ is quite small compared to the solid clusters.

\begin{table}
\caption{\label{restitution} Simulation data of coefficients of restitution $e$ with impact velocity v.}
 \begin{tabular}{||c|c|c|c|c|c|c||} \hline
 v & 0.1 & 0.2 & 0.5 & 1.0 & 1.5 & 2.0 \\\hline
 e & 0.437 & 0.349 & 0.259 & 0.174 & 0.131 & 0.105 \\\hline
 \end{tabular}
\end{table}

Now, let us describe the overall collision processes. The collision snapshots are shown in Fig. /ref{overall} at four different times at t = 0, 5.0, 27.5, and 75.0. Of course, the shapes of colliding droplets significantly change with time as expected. At $t = 0.0$, one droplet has roughly a spherical shape with a radius $R \approx 22.09$ and the separation gap between two droplets is set at $\Delta x_0=5.0$. They are approaching each other with a relative velocity $2u_o=3.0$. The spheres are unilaterally compressed along the longitudinal direction up to $t = 6.5$, as shown in Fig. /ref{overall}b). Note that the colliding region is squeezed into a flat surface, but the rest still keeps a spherical shape. Then the droplet is elongated along the transverse directions (in y-z plane) up to $t = 30.0$ as shown in Fig. /ref{overall}c). At around $t = 32.0$, its shape is maximally deformed; it becomes roughly a prolate ellipsoid with a length of minor axis $L_x \approx 18.15$ and lengths of major axis $L_y \approx 69.57$ and $L_z \approx 71.22$. Then follows a recovery returning to its starting shape up to $t = 55.0$. At $t = 55.0$, the droplets are finally separated and begin to recoil each other, \textit{i.e.} they are eventually receding with a recoil velocity $v'_x\approx0.14$ with some small acquired transverse velocities. Therefore, they are moving at an odd angle from each other, even though the collision is assumed to be head-on. During the recoil, they may tend to return to a spherical shape due to their surface tension. In other words, they are in the stress relaxation with a long relaxation time.

Next, to examine the elastic behavior of the collision process, we focus on the first part of initial compressed phase. To see more detailed motion, we halved the size of timestep during initial compressed phase, since the particles involved during the collision are moving rather rapidly in this phase. In Fig. \ref{f4}, we present the force $F_n$ \textit{vs}. deformation $\xi$ in the initial compressed phase up to $t=5.0$. The deformation $\xi$ can be directly measured with the standard image analysis. The dotted points are the simulation results. The solid line is fitted to the Eq. (\ref{deformation}). This may come from two reasons: first, the very initial contact tip is not a continuous media. They are composed of only a small number of molecules. Second, the shape of the tip is not spherical, but irregular and rough. The fitting is rather good except at small values of $F_n$. From the fitting to the data using Eq. (\ref{deformation}), we obtain a value of
\begin{equation}
    (2\gamma^2 / R)^{1/3} = 0.0253. \label{xifit}
\end{equation}

\begin{figure}[htbp]
\begin{center}
\includegraphics[scale=.6]{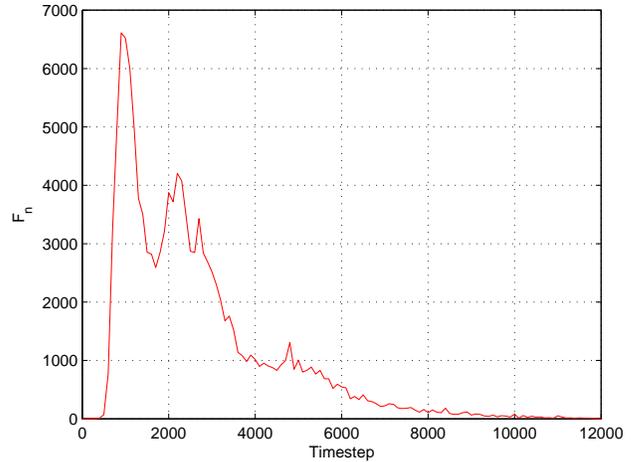}
\caption{(Color on-line) Overall longitudinal force $F_x$ \textit{vs.} time t. The droplets are approaching to each other with a velocity $u_0 = 1.5$. Note the second peak around at t = 11.0.}
\label{f3}
\end{center}
\end{figure}

\begin{figure}
\centering
\begin{tabular}{c c}
a) \includegraphics [scale=.15]{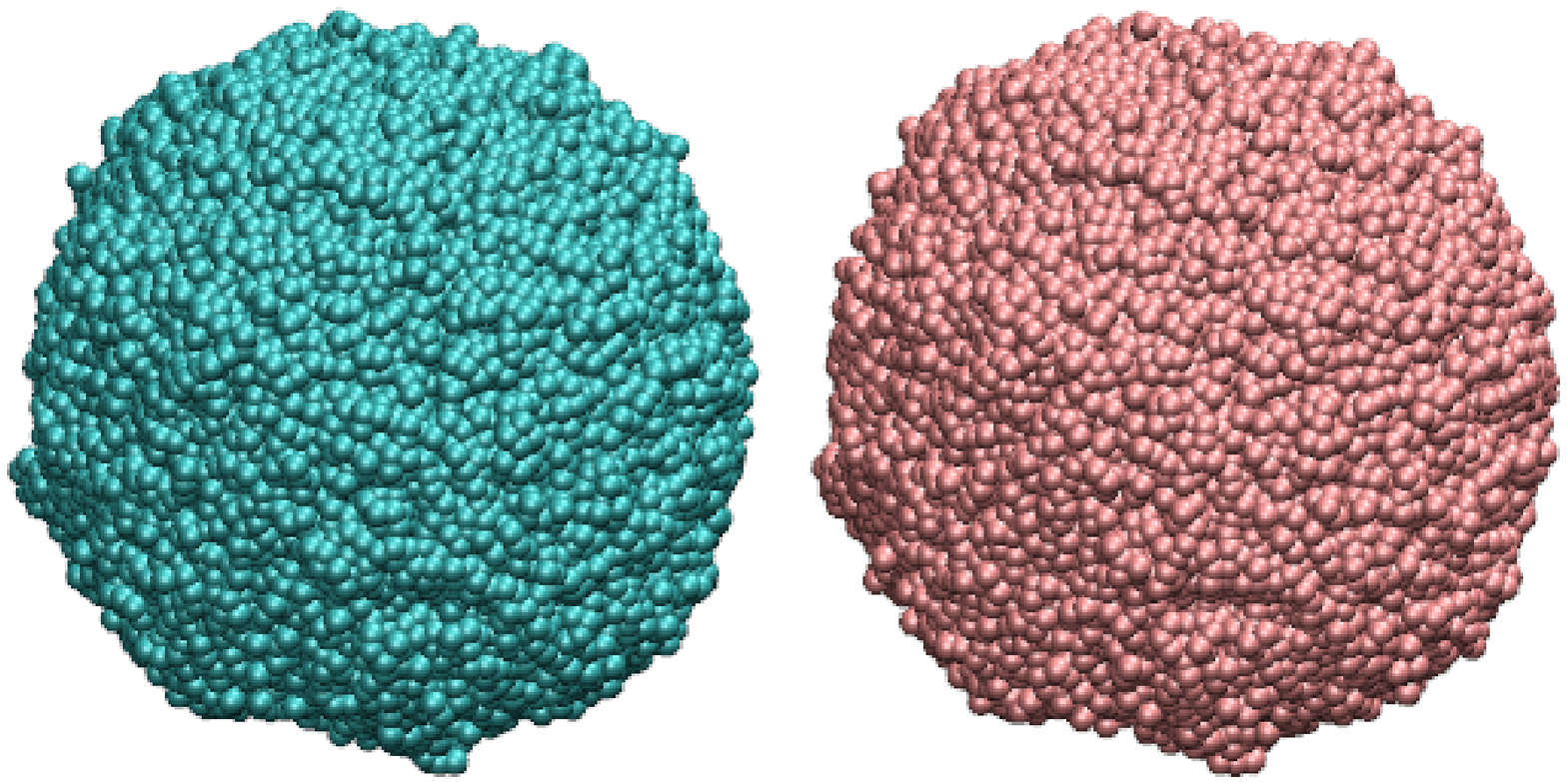}&
b) \includegraphics [scale=.15]{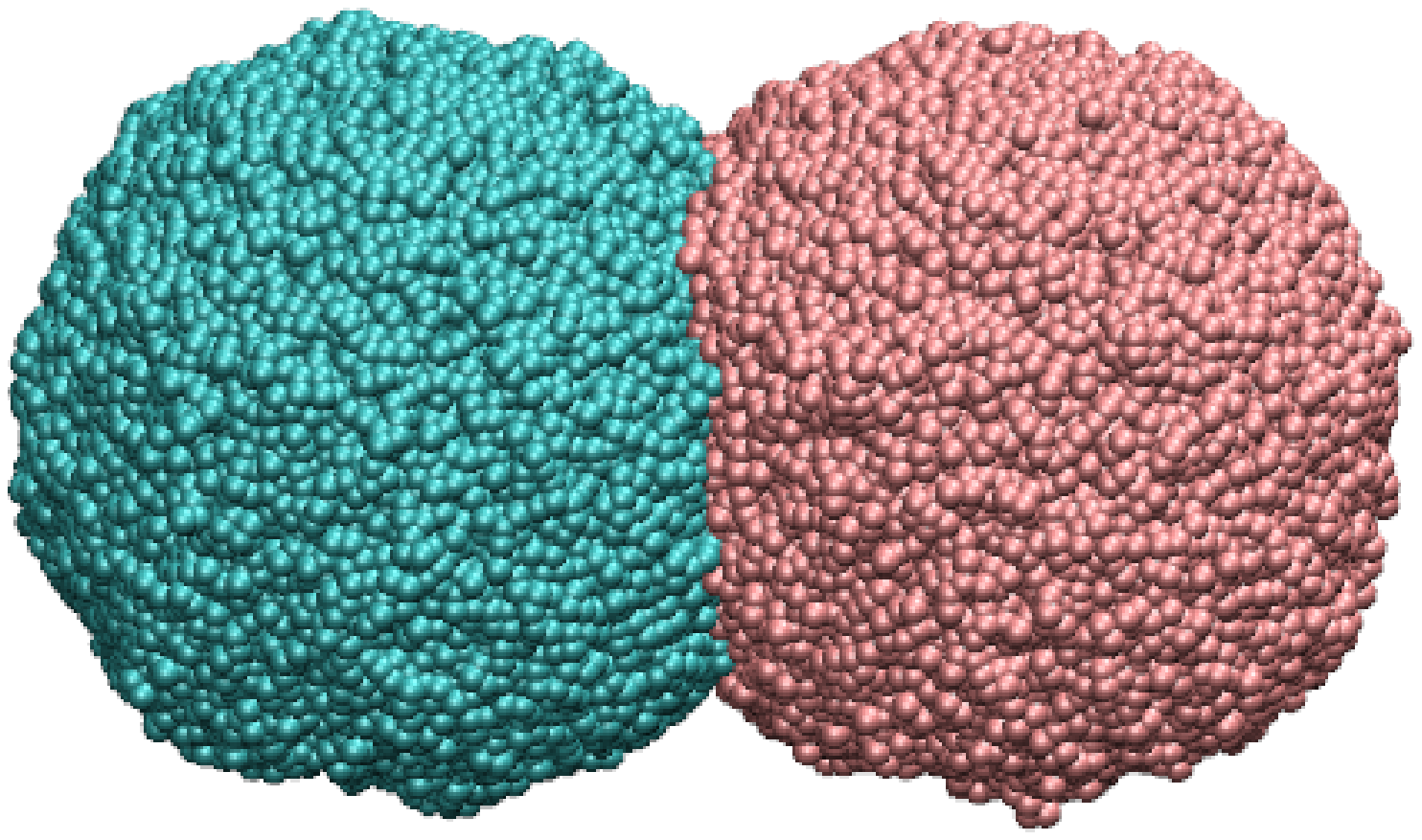}  \\
c) \includegraphics [scale=.15]{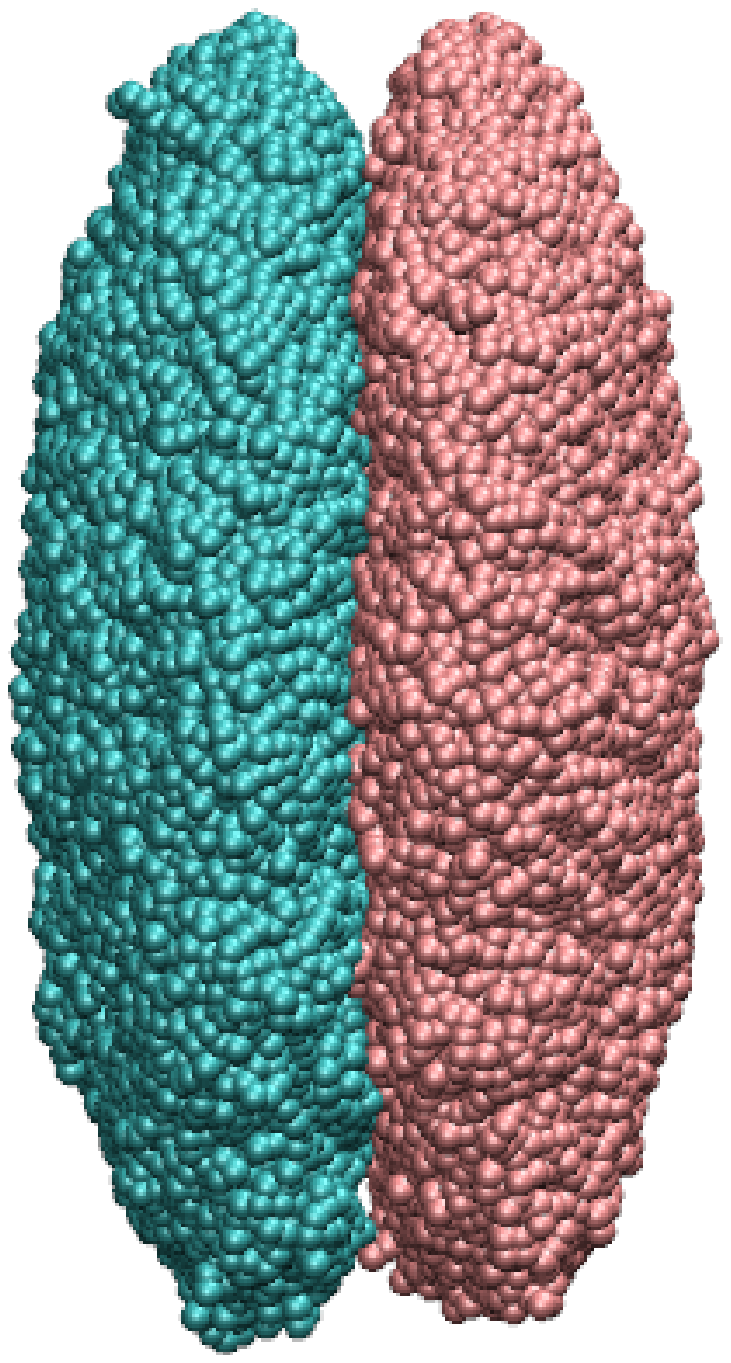}  &
d) \includegraphics [scale=.15]{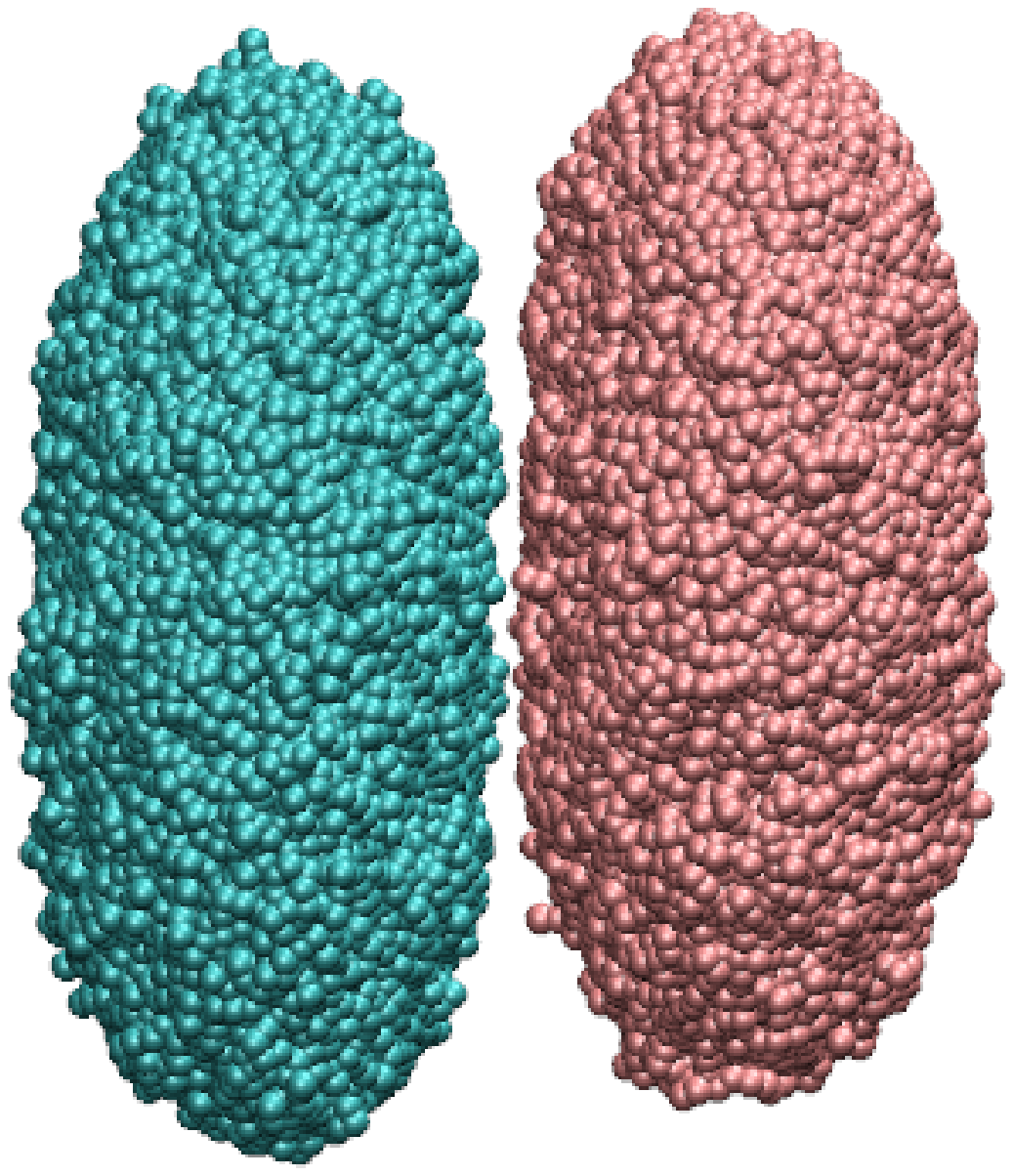} \\
\end{tabular}
\caption{(Color on-line) Snapshots of collision progresses at different times. a) t = 0, separated with gap of $\Delta x_0=5.0$ between two leading edges, approaching with a impact velocity $2u_o=3.0$, b) t = 5.0, initially compressing stage, still maintaining a spherical shape, c) t = 27.5, transversely expanding, prolate ellipsoidal shape, d) t = 75.0, receding stage, in the process of returning to its spherical shape.}
\label{overall}
\end{figure}

\begin{figure}[htbp]
\begin{center}
\includegraphics[scale=.6]{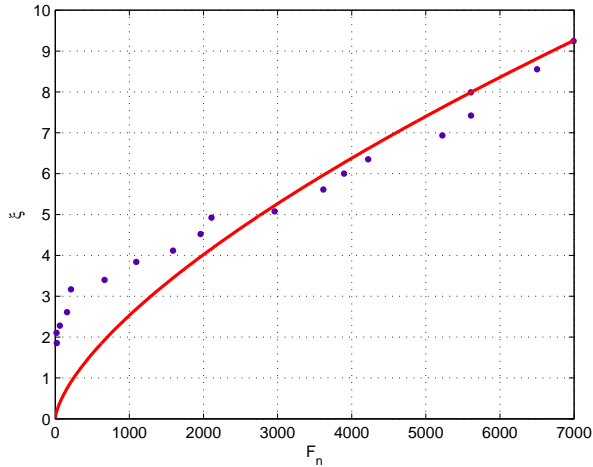}
\caption{(Color on-line) Deformations $\xi$ \textit{vs.} longitudinal force $F_n$ in the initial compressed phase of a collision. Solid line is a fit to the data by Eq. (\ref{deformation}) with $R^2 = 0.73$ and $({\gamma^2 2/R})^{1/3} = 0.0253$. Note the poor agreement near the origin.}
\label{f4}
\end{center}
\end{figure}

\begin{figure}[htbp]
\begin{center}
\includegraphics[scale=.6]{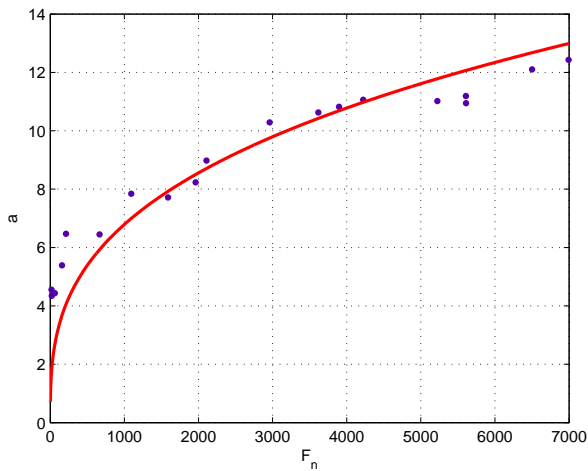}
\caption{(Color on-line) Contact radius $a$ \textit{vs.} longitudinal force $F_n$ in the initial compressed phase of a collision. Solid line is a fit to the data by Eq. (\ref{c_radius}) with $R^2 = 0.78$ and $({\gamma R/2})^{1/3} = 0.679$. Note again the poor agreement near the origin.}
\label{f5}
\end{center}
\end{figure}

Finally, to further confirm the validity of the theoretical prediction for the elastic theory of the initial compressed phase of a droplet, the contact radius $a$ in Eq. (\ref{c_radius}) is calculated. The consequent results of the contact radius $a$ \textit{vs.} force $F_n$ are depicted in Fig. \ref{f5}. The values are also measured with the standard image analysis. In the same time range, the deformation $\xi$ \textit{vs.} force $F_n$ in  Eq. \ref{f4} show the similar behavior as the prediction of Eq. (\ref{c_radius}). In Fig. \ref{f5}, when the radius $a$ is smaller than 6.0, the force $F_n$ is shapely increasing. Or, at around $F_n = 800$, a sharp increase of stretched S-shape is observed. But at around $a = 10.0$, they show a linear increase. From the fitting to the data using Eq. (\ref{c_radius}), we can again obtain a value of
\begin{equation}
    (\gamma R/2)^{1/3} = 0.679. \label{afit}
\end{equation}

From the Eqs. (\ref{xifit}) and (\ref{afit}), we can get the estimated value of $R_{est} \approx 36.45$, which is rather larger than our initially setting value of $R = 22.87$, but the order of magnitude is comparable.

\section{Conclusion}
Let us now conclude our results. Here, we mainly focused on the initial compressed phase for the head-on collision of equal-sized spherical polymer nanodroplets. In this phase, the phenomenon can be explained by an elastic compressed phase in longitudinal direction. When the colliding region is too compressed to keep the spherical shape of the droplet, its shape changes into an prolate ellipsoid, as shown in Fig. 3c). The next phase is thus expansion in the transverse direction. After the compression, energy exceeds the elastic limit to support the spherical shape, it may then break into parts. The elastic behaviors of the initial compressed phase of the collision can be understood with the Hertzian theory of elastic balls. The simulation results conform to this theoretical predictions. At least, the initial compressed phase of a collision, the Hertzian theory of elastic balls explains rather well the simulation results as measured with deformations $\xi$, longitudinal forces $F_n$, and contact radii $a$, except the very moment of collision. In this limit, the continuum theory of the Hertz model is broken, since there are very small number of molecules involved in the collision and furthermore, the shape of the initial contact tip is far from spherical, but irregular and rough.

More MD simulations should be followed at different thermodynamic states and simulation control parameters for further confirmations. We considered only the initial compressed phase in this paper. To explain properly the later phases of the collision process, we need also a more elaborate theory of viscoplasticity or viscoelasticity.


\end{document}